\documentstyle[12pt]{article}
\begin{document}
\begin{center}
{\large\bf The Exact Solution of one Fokker-Planck Type Equation }

{\large\bf used by R.Friedrich and J.Peinke}

{\large\bf in the Stohastic Model of a Turbulent Cascade}

\vspace*{0.15truein}
{A. A. Donkov$^1$, A. D. Donkov$^2$ and E. I. Grancharova$^3$}
\vspace*{0.15truein}

$^1${\it Dept.of Physics, University of Wisconsin, 1150 Univ-Ave,
Madison, WI-53706, USA}

e-mail : donkov@phys-next1.physics.wisc.

$^{2}${\it Dept.of Physics, University of Sofia,
 5 J.Bourchier Blvd., Sofia 1164, BULGARIA}

e-mail: donkov@phys.uni-sofia.bg \\
and {\it Bogoliubov Laboratory of Theoretical Physics, JINR, Dubna,
Moscow Region, RUSSIA}\\
e-mail: donkov@thsun1.jinr.ru

$^{3}${\it Dept.of Physics, University of Sofia,
 5 J.Bourchier Blvd., Sofia 1164, BULGARIA}\\

e-mail: granch@phys.uni-sofia.bg
\end{center}
\vspace*{0.15truein}
\begin{abstract}
The exact solution of the Cauchy problem for a Fokker-Planck equation
used by R.Friedrich and J.Peinke for the description of a turbulent
cascade, considered as a stochastic process of Markovian type,
is obtained in the frame of M.Suzuki approach.
\end{abstract}

\section{Introduction}
The understanding of the turbulence is one of the main unsolved problems
of classical physics, in spite of the more than 250 years of
strong investigations initiated by D.Bernoulli and L.Euler.

In the stochastic approach to turbulence~\cite{Monin},~\cite{Frish}
the turbulent cascade is
considered as a stochastic process, described by the probability
distribution $P(\lambda,v)$, where $\lambda$ and $v$ are the
appropriate scaled length and velocity increment respectively.
Recently~\cite{Friedrich} R.Friedrich and J.Peinke presented
experimental evidence that 
the probability
density function  $P(\lambda,v)$ obeys a Fokker-Planck
equation (FPE)~\cite{Risken} (see fig.1 and fig.2
in~\cite{Friedrich}):
\begin{equation}
\frac{\partial P(\lambda,v)}{\partial \lambda} =
\left[ -\frac{\partial}{\partial v} D^1(\lambda,v)
+ \frac{\partial^2}{\partial v^2} D^2(\lambda,v) \right]
P(\lambda,v),
\label{FPP}
\end{equation}
where the drift and duffusion coefficients
$D^1$ and $D^2$ respectively are derived by analysis of experimental
data of a fluid dynamical experiment (see fig.3 in~\cite{Friedrich}).

In their paper Friedrich and Peinke consider the application
of the FPE to obtain the Kolmogorov scaling  with simplified
assumptions that $D^1$ and $D^2$ are $\lambda$-independent,
$D^1$ is linear in $v$ and $D^2$ is quadratic in $v$:
$$
D^1= -a\, v, \qquad a>0; \qquad
D^2 = c\, v^2,\qquad c>0 \,.
$$
( In the notations of~\cite{Friedrich} :
$ a \equiv \gamma $ and $ c \equiv Q $.)

Here we will consider a more realistic situation
(see fig.3 in~\cite{Friedrich})
of $\lambda$-dependent $D^1$ and $D^2$ :
\begin{equation}
D^1= -a(\lambda)\, v, \qquad a(\lambda)>0 ;\qquad
D^2 = c(\lambda)\, v^2,\qquad c(\lambda)>0 .
\label{D}
\end{equation}
Thus the FPE ~(\ref{FPP}) will take the form:
\begin{equation}
\frac{\partial P}{\partial \lambda} = b_0(\lambda)P(\lambda,v)+
b_1(\lambda) v \frac{\partial P}{\partial v} +
c(\lambda)\left( v \frac{\partial}{\partial v}\right)^2 P(\lambda,v),
\label{P}
\end{equation}
where
\begin{equation}
b_0(\lambda)= a(\lambda)+2 c(\lambda),
\quad b_1(\lambda) = a(\lambda) + 3 c(\lambda) .
\label{b}
\end{equation}

\begin{center}
\section{Exact Solution of the Cauchy Problem for the
Eq.~(\ref{P})}
\end{center}
In this section we will find the solution
$P(\lambda,v)$
 of the Cauchy problem
for the Eq.~(\ref{P}) with the initial condition
\begin{equation}
P(0,v) = \varphi (v).
\label{In}
\end{equation}
According to \cite{Monin}$-\!$~\cite{Friedrich}, when the probability
density function is known, one may derive all properties
of the turbulent cascade considered as a stochastic process.

For the solution of the problem (\ref{P}),~(\ref{In}) we
shall use the approach
of M.Suzuki~\cite{Suzuki} to the FPE (see also ~\cite {Donkov} ),
based on the disentangling techniques of R.Feynman~\cite{Feynman}
 and the operational methods developed in the functional
analysis, in particular in the theory of pseudodifferential equations
with partial derivatives ~\cite{Hoermander}$-\!$~\cite{Maslov}

In the spirit of the operational methods using the
pseudodifferential operators  we can write the solution
of the Cauchy problem (\ref{P}),~(\ref{In}) in the form
\begin{equation}
P(\lambda, v) =
\left(exp_+ \int_0^{\lambda} \left[ b_0(s)+b_1(s)v
\frac{\partial}{\partial v}
+ c(s)\left(v \frac{\partial}{\partial v}\right)^2 \right] {\rm d}s
\right) \varphi(v) ,
\label{form}
\end{equation}
where the symbol $\;\;exp_+\;\;$ designates the V.Volterra ordered exponential
\begin{equation}
exp_+ \int_0^{\lambda} \hat C(s) {\rm d}s =
\hat 1 + \lim_{n\to\infty} \sum_{k=1}^n\int_0^{\lambda}{\rm d}\lambda_1
\int_0^{\lambda_1}{\rm d}\lambda_2 \dots \int_0^{\lambda_{k-1}}
{\rm d}\lambda_{k}
\hat C(\lambda_1) \hat C(\lambda_2) \dots \hat C(\lambda_{k}).
\label{exp}
\end{equation}

The linearity of the integral and the explicit form of the operators
in~(\ref{form}) permit to write the solution $P(\lambda,v)$ in terms
of usual, not ordered, operator valued exponent
\begin{equation}
P(\lambda,v) = {\rm e}^{\beta_0 (\lambda)}\,
{\rm e}^{\beta_1(\lambda)v\frac{\partial}{\partial v} +
\gamma (\lambda)\left(v\frac{\partial}{\partial v}\right)^2}
\varphi(v) ,
\label{expP}
\end{equation}
where for convenience we have denoted
\begin{equation}
\beta_j(\lambda) = \int_0^{\lambda}b_j(s){\rm d}s,\;\; (j=0,1);
\qquad \gamma(\lambda) = \int_0^{\lambda}c(s) {\rm d}s .
\label{beta}
\end{equation}
Consequently (from now on "$'$" means $\frac{\rm d}{{\rm d}t} $ ) :
\begin{equation}
\beta_j(0)=0,\;\;\; {\beta}'_j(\lambda) =b_j(\lambda),\;\;\; (j=0,1);
\qquad \gamma(0)=0,\;\;\; {\gamma}'(\lambda)=c(\lambda).
\label{betaPR}
\end{equation}

Since the operators \qquad
$\hat A \equiv \beta_1(\lambda) v \frac{\partial}{\partial v}$ \quad and \quad
$\hat B \equiv \gamma (\lambda)\left(v \frac{\partial}{\partial v}\right)^2$\qquad
commute : $[\hat A , \,\hat B] = 0$ ,
from Eq.~(\ref{expP}) we have

\begin{equation}
P(\lambda,v)=
{\rm e}^{\beta_0(\lambda)}\,
{\rm e}^{\beta_1(\lambda) v\frac{\partial}{\partial v}} \,
{\rm e}^{\gamma(\lambda)\left(v\frac{\partial}{\partial v}\right)^2}
\varphi(v).
\label{Form}
\end{equation}

Therefore, taking into account the formulae
\begin{equation}
{\rm e}^{\beta_1(\lambda)v\frac{\partial}{\partial v}}f(v)
= f\left(v{\rm e}^{\beta_1(\lambda)}\right)
\label{F1}
\end{equation}
and
$$
\,\,{\rm e}^{\gamma(\lambda)\left(v\frac{\partial}{\partial v}\right)^2}g(v)
=\frac{1}{\sqrt{4\pi\gamma(\lambda)}}\int_{-\infty}^{\infty}
{\rm e}^{-\frac{s^2}{4\gamma(\lambda)}} g\left(v{\rm e}^
{-s}\right){\rm d}s
$$
\begin{equation}
\qquad\qquad\qquad=\frac{1}{\sqrt{4\pi\gamma(\lambda)}}
\int_{-\infty}^{\infty}
{\rm e}^{-\frac{\left(\ln v-y\right)^2}{4\gamma(\lambda)}}
g\left({\rm e}^
{y}\right){\rm d}y,
\label{F2}
\end{equation}
we obtain the following expression for the exact solution of the
Cauchy problem (\ref{P}),(\ref{In})
$$ 
P(\lambda,v)
=\frac{{\rm e}^{\beta_0(\lambda)}}
{\sqrt{4\pi\gamma(\lambda)}}\int_{-\infty}^{\infty}
{\rm e}^{-\frac{s^2}{4\gamma(\lambda)}} \varphi\left(v{\rm e}^
{\beta_1(\lambda)-s}\right){\rm d}s
$$
\begin{equation}
\label{end}
\qquad\qquad=\frac{{\rm e}^{\beta_0(\lambda)}}
{\sqrt{4\pi\gamma(\lambda)}}\int_{-\infty}^{\infty}
{\rm e}^{-\frac{\left(\ln v+\beta_1(\lambda)-y\right)^2}
{4\gamma(\lambda)}}g\left({\rm e}^{y}\right){\rm d}y,
\end{equation}
where $\beta_0(\lambda), \beta_1(\lambda)$  and $\gamma(\lambda)$
are defined in~(\ref{beta}).

Substituting the expression~(\ref{end})
in the Eqs.~(\ref{P}) and~(\ref{In}) and using the Eq.~(\ref{betaPR})
one can see immediately that $P(\lambda,v)$ is a solution of the
problem (\ref{P}),~(\ref{In}) and, according to the Cauchy theorem,
it is the only classical solution of this problem.

\section{Concluding remarks}
\begin{itemize}
\item The exact solution of the Cauchy problem (\ref{P}),~(\ref{In})
is obtained using the algebraic method we have described.
The Eq.~(\ref{P}) is a generalization
of the equation used by R.Friedrich and J.Peinke
( see section 1)
in their description of a turbulent cascade by a
Fokker-Planck equation
with coefficients derived by a detailed analysis of
experimental data
of a fluid dynamical experiment.
\item If 
the probability distribution function
$P(\lambda,v)$ is known, then
one may derive the properties of
a given stochastic process, in our case -
 the turbulent cascade \cite{Monin}~$-\!$~\cite{Friedrich}.
\item For more realistic description of the turbulent cascade
by a FPE
it should be desirable to use for
$D^1(\lambda, v)$ and $D^2(\lambda ,v)$ in the Eq.~(\ref{FPP})
more general expressions than these in Eq.~(\ref{D}),
for instance:\\
 $ D^1(\lambda,v) = a_1(\lambda) - a(\lambda)v,\,\,\, a(\lambda)>0$\quad
and \quad
$D^2(\lambda,v) = c_1(\lambda) + c(\lambda) v^2 $.

\end{itemize}

\end{document}